\documentclass[apj]{emulateapj}

\newcommand{\hi}          {\mbox{\rm \ion{H}{1}}}
\newcommand{\hii}         {\mbox{\rm \ion{H}{2}}}

\newcommand{\kms}         {km~s$^{-1}$}

\newcommand{\mlk}         {\mbox{$M_{\odot}/L_{\odot K}$}}

\newcommand{\mslk}        {\mbox{$M_{*}/L_{K}$}}

\newcommand{\hr}          {\mbox{$^h$}}
\renewcommand{\min}       {\mbox{$^m$}}

\shorttitle{Is There a Fundamental Line for Disk Galaxies?} 
\shortauthors{Simon et al.}
\slugcomment{Accepted for publication in \emph{The Astrophysical Journal}}

\begin{document}

\title{Is There a Fundamental Line for Disk Galaxies?}

\author{Joshua D. Simon\altaffilmark{1}, Francisco
Prada\altaffilmark{2}, Jos\'{e} M. V\'{\i}lchez\altaffilmark{2}, 
Leo Blitz\altaffilmark{3}, and Brant Robertson\altaffilmark{4}}

\altaffiltext{1}{Department of Astronomy, California Institute of Technology
                 1200 E. California Blvd, MS 105-24, Pasadena, CA  91125;
                 jsimon@astro.caltech.edu}

\altaffiltext{2}{Instituto de Astrof\'{\i}sica de Andaluc\'{\i}a, CSIC
                 Apdo. 3004, 18080 Granada, Spain;
	         fprada@iaa.es, jvm@iaa.es} 

\altaffiltext{3}{Department of Astronomy, University of California at Berkeley
                601 Campbell Hall, Berkeley, CA  94720;
	        blitz@astro.berkeley.edu}

\altaffiltext{4}{Harvard-Smithsonian Center for Astrophysics, 
                60 Garden Street, Cambridge, MA  02138;
	        brobertson@cfa.harvard.edu}

\begin{abstract}

We show that there are strong local correlations between metallicity,
surface brightness, and dynamical mass-to-light ratio within M33,
analogous to the fundamental line of dwarf galaxies identified by
\citet{pb02}.  Using near-infrared imaging from 2MASS, the published
rotation curve of M33, and literature measurements of the
metallicities of \hii\ regions and supergiant stars, we demonstrate
that these correlations hold for points at radial distances between
140~pc and 6.2~kpc from the center of the galaxy.  At a given
metallicity or surface brightness, M33 has a mass-to-light ratio
approximately four times as large as the Local Group dwarf galaxies;
other than this constant offset, we see broad agreement between the
M33 and dwarf galaxy data.  We use analytical arguments to show that
at least two of the three fundamental line correlations are basic
properties of disk galaxies that can be derived from very general
assumptions.  We investigate the effect of supernova feedback on the
fundamental line with numerical models and conclude that while
feedback clearly controls the scatter in the fundamental line, it is
not needed to create the fundamental line itself, in agreement with
our analytical calculations.  We also compare the M33 data with
measurements of a simulated disk galaxy, finding that the simulation
reproduces the trends in the data correctly and matches the
fundamental line, although the metallicity of the simulated galaxy is
too high, and the surface brightness is lower than that of M33.

\end{abstract}

\keywords{galaxies: abundances --- galaxies: formation --- galaxies:
fundamental parameters --- galaxies: individual (M33) --- galaxies:
kinematics and dynamics --- galaxies: spiral}

\section{INTRODUCTION}
\label{introduction}

\citet[][hereafter PB02]{pb02} recently found that the dwarf galaxies
in the Local Group obey a puzzling correlation between mean
metallicity and global dynamical mass-to-light ratio (M/L).  By
combining this relationship with the previously known trend of
metallicity with central surface brightness \citep[e.g,][]{ped90,c98},
PB02 demonstrated that all three parameters together form an even
stronger relationship, which they labeled the fundamental line of
dwarf galaxies.  PB02 showed that the metallicity-M/L ratio
correlation can be explained by a simple chemical enrichment model.
In this model, galactic winds continuously remove the heavy elements
formed in dwarf galaxies from their interstellar medium, and star
formation in these galaxies ends when all of the gas has been blown
out of the system.  Because more metals are retained by more massive
galaxies (which have lower mass-to-light ratios) this model is able to
reproduce the observed metallicity-M/L relationship.  However, the
model does not make clear why including surface brightness as a third
parameter improves the correlation, and it is also surprising that
variables such as morphological type and star formation history have
no apparent effect on the fundamental line.  PB02 proposed that
feedback from star formation or supernovae could be involved in
regulating the fundamental line, and \citet{dw03} and \citet{tassis03}
used analytical arguments and simulations, respectively, to show that
supernova feedback can produce relationships like the fundamental
line, but additional work is needed to improve this understanding.

Two natural questions arise from these previous investigations of the
fundamental line.  First, does the fundamental line hold for massive
galaxies as well as dwarfs?  Galactic winds and supernova feedback
should operate less effectively in larger galaxies as a result of
their deeper potential wells, so if either of these processes is the
primary mechanism responsible for setting the fundamental line
correlations, the fundamental line should not be the same for massive
systems.  Second, is the fundamental line a local law as well as a
global one?  If not, it would be puzzling that all dwarf galaxies know
to follow the relationship globally without having any corresponding
correlations on smaller scales.  The data to test the fundamental line
locally in dwarfs do not yet exist, because spatially resolved
metallicity measurements have not been obtained, and few dSph and dE
galaxies have well-resolved kinematics either.  Answering either of
these questions with existing data therefore requires us to move to
more massive galaxies.

In order to explore further the nature of the fundamental line, we
consider the case of a well-resolved disk galaxy with known variations
of its metallicity, surface brightness, and mass-to-light ratio with
radius.  The best example of a galaxy with these necessary
characteristics is M33.  We will address the following questions with
this study.  1) Does the fundamental line hold locally, at every point
within a galaxy?  2) Do the parameters of the global and local
(assuming one exists) fundamental lines agree?  3) What is the origin
of the fundamental line?  4) Can simulations of galaxy formation
reproduce the fundamental line?

In this paper, we use several recently published data sets to
investigate the local fundamental line in M33.  In the next section,
we combine a deep Two Micron All Sky Survey (2MASS) mosaic of M33
\citep{block04}, high resolution CO and \hi\ rotation curves
\citep{c03,cs00}, and literature metallicity measurements to construct
the necessary observational quantities.  In \S \ref{relationships} we
perform fits to the various relationships between mass-to-light ratio,
metallicity, and surface brightness, and in \S \ref{discussion} we
compare the locally determined fundamental line of disk galaxies to
the fundamental line of dwarf galaxies, give derivations of two of the
fundamental line relationships, and discuss the implications of our
findings.  We also use the recent $\Lambda$CDM disk galaxy simulation
by \citet{robertson04} to test how accurately current simulations can
reproduce the properties of galaxies and we employ additional toy
models of galaxy formation to understand the impact of feedback on the
fundamental line.  We briefly summarize our results and conclude in \S
\ref{conclusions}.

\section{DATA AND ANALYSIS}
\label{observations}

Of the observations required for this study, the most restrictive is
the large sample of accurate metallicity measurements as a function of
radius.  Many nearby galaxies have high-quality rotation curves and
deep optical imaging, although $K_{s}$ band imaging to the depth of
the 2MASS images (see \S \ref{2mass}) is less common.  However, only
two galaxies --- M101 \citep{kbg03} and M33 --- have measured
abundances for more than $\sim10$ \hii\ regions.  Because M101 is a
massive galaxy (and therefore more dynamically dominated by baryons),
is nearly face-on (making its kinematics more difficult to measure
accurately), and is significantly lopsided (indicating a possible
recent interaction), it is not an ideal target for this study, and we
instead focus exclusively on M33.

\subsection{Abundances}
\label{abundances}

Studying the fundamental line over a large range of galactic
environments requires abundance measurements covering as wide a range
of radii as possible.  Because we will be comparing abundances to
other local properties, it is important that we use a consistent
abundance scale.  We therefore limit our \hii\ region sample to
objects with directly-determined electron temperatures (although we do
include CC93, the closest \hii\ region to the center of M33, for which
abundances were derived from strong-line model fitting by
\citealt{v88}).  We also utilize abundances for M33 B supergiants,
which have metallicities consistent with those of the \hii\ regions in
M33, although with slightly larger scatter.

\begin{figure}[t!]
\epsscale{1.20}
\plotone{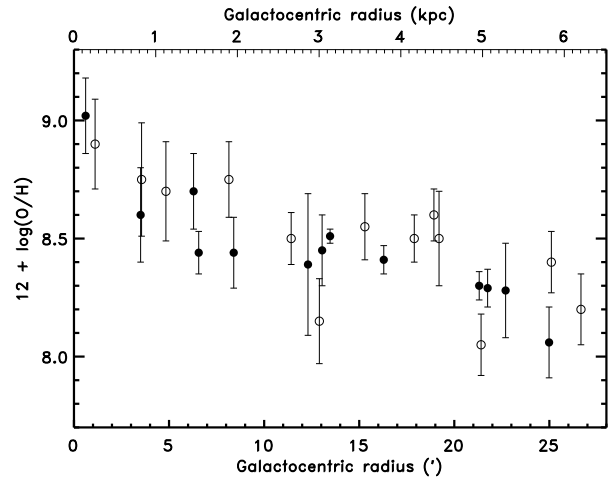}
\caption{Oxygen abundances as a function of radius in M33.  The filled
  points represent abundance measurements from \hii\ regions and the
  open points represent abundance measurements from B supergiants. }
\label{metalfig}
\end{figure}

The abundance of oxygen in the interstellar medium (ISM) and in the
young stellar population of M33 has been sampled across the disk of
the galaxy using both star-forming regions and young stars.  We have
compiled abundance data for 13 \hii\ regions derived from optical
spectroscopy \citep{smith75,ka81,v88} and ISO LWS spectroscopy
\citep{higdon}.  The compilation of abundances for 13 bright M33
supergiants came from \citet{monteverde97}, \citet{u04}, and
\citet{u05}.  These works used quantitative spectroscopy of M33
supergiants, subsequently analysed with state of the art model
atmospheres of massive stars.  Combining these data sets yields 26
metallicity measurements in M33, spanning radii from 140~pc out to
6.2~kpc.  The metallicity measurements employed in our study are
plotted as a function of radius in Figure \ref{metalfig}.  Note that
we assume a distance of 800~kpc throughout this paper
\citep{lee02,alan04}, although some other recent determinations
suggest a distance of more than 900~kpc \citep{kim02,ciardullo04}.

Abundance data for other astrophysical sources have been reported in
the literature for M33: e.g., planetary nebulae
\citep[PN;][]{magrini03,magrini04} and supernova remnants
\citep[SNR;][]{bk85}; overall the PN and SNR abundances show a general
agreement with the abundances derived for the \hii\ regions
\citep[see][]{v88,magrini04,stasinska05}.  However, only small samples
of these objects have been observed, and the scatter in the data is
large, indicating that further observations are needed to understand
how accurately they trace the present ISM abundances.  Therefore, for
the purposes of this paper we have not used PN and SNR data to trace
the abundance gradient, though we point out that PN and (possibly
also) SNR abundance results appear consistent with the data set used
here.

\subsection{2MASS Images}
\label{2mass}

In addition to providing relatively shallow near-infrared images of
the entire sky, the 2MASS project also made deeper observations of
several nearby galaxies.  The deep exposures of M33 have integration
times six times longer than the standard 2MASS data, so they reach
$\sim1$ mag fainter in surface brightness \citep{block04}.  Tom
Jarrett has kindly provided us with a mosaic of the individual images
that covers the entire galaxy, with foreground stars and background
(sky) variations removed.

$K_{s}$ band surface brightnesses at the position of each \hii\ region
and B star were measured by adding up the flux contained in circular
apertures of radius 7\arcsec.  The local surface brightnesses measured
in this way are consistent with the azimuthally averaged surface
brightness (see below) at the same radius within the uncertainties.
Although the \hii\ regions themselves contribute negligibly to the
observed flux at this wavelength, in some cases a single star
coincidentally located very close to the \hii\ regions adds
significantly to the flux, biasing the surface brightness high.  To
avoid this contamination, in such cases we instead measured the
surface brightness in a blank region of sky a few arcseconds away.

\begin{figure}[t!]
\epsscale{1.20}
\plotone{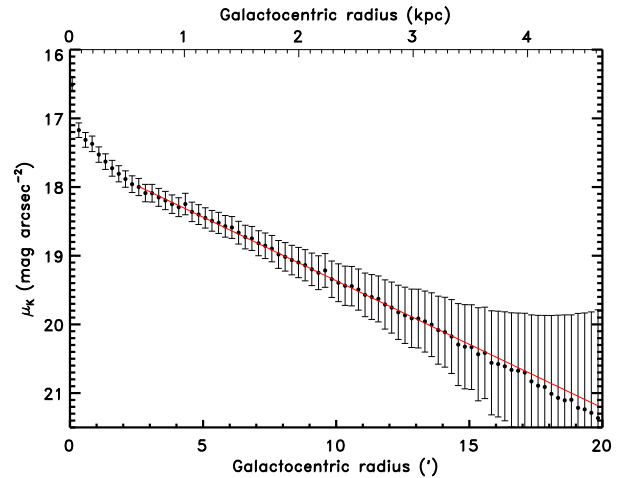}
\caption{$K$-band surface brightness profile of M33 from 2MASS data.  
The error bars represent formal statistical uncertainties from the
isophotal fits, which clearly overestimate the true uncertainties.
The solid red line shows the exponential disk fit to the surface
brightness profile.}
\label{sbfig}
\end{figure}

To obtain deprojected galactocentric distances of the \hii\ regions
and B stars, we assumed a constant position angle of $22\degr$ and a
constant inclination angle of $50\degr$ (see \S \ref{rc}).  The
ellipses were centered on the galaxy nucleus at $01\hr33\min50\fs90$
$+30\degr39\arcmin36\farcs2$, which is consistent with the listed
position of M33 in the 2MASS Extended Source Catalog (see also
\citealp{lga}).  The surface brightness profile that we derive is
displayed in Figure \ref{sbfig} along with an exponential disk fit.
The large-scale background subtraction that was performed on the image
makes accurate assessment of surface brightness errors extremely
difficult.  We therefore added a minimum uncertainty of
0.1~mag~arcsec$^{-2}$ in quadrature to the measured Poisson
uncertainties (T. Jarrett 2004, personal communication).  Our fits to
the surface brightness profile (see \S \ref{rc}) yielded an
extrapolated central surface brightness for the disk of $\mu_{K} =
17.51 \pm 0.06$ and a disk scale length of $5.86\arcmin \pm
0.28\arcmin$ (1.36~kpc).  This value for the scale length is in
agreement with the previous near-infrared photometric analysis by
\citet{rv94}, but the central surface brightness we measure is
$\sim0.15$~mag brighter, even after accounting for the different
inclination angles.  Because M33 is very extended compared to the
largest near-infrared CCD mosaics currently available, determining an
accurate photometric zeropoint for the surface photometry is quite
difficult, which is likely the cause of the different central surface
brightnesses derived by us and \citet{rv94}.  The imaging by
\citet{rv94} did not reach the edge of the galaxy, suggesting that
they may have derived an incorrect sky level, but the sky subtraction
is sufficiently problematic that either or both works may have a
systematic zeropoint offset of $\sim0.15$~mag.

\subsection{Rotation Curve}
\label{rc}

In order to calculate the dynamical mass of M33, we need a rotation
curve.  The best existing rotation curves of M33 are those constructed
by \citet{c03} in CO and \citet{cs00} in \hi, which are both
consistent with the high-resolution CO observations of \citet{greg}.
\citet{cs00} fit a tilted-ring model to the \hi\ velocity field that
they observed with the Arecibo telescope.  They augmented their low
resolution (beam FWHM of 3\farcm5 = 790 pc) data with interferometric
observations near the center of the galaxy by \citet{newton80} with a
beam size of $47 \times 93\arcsec$.  The best-fitting model
constructed by both sets of authors has a position angle (PA) of
$\sim22\degr$ and an inclination angle of $\sim50\degr$ over the
portion of the galaxy covered by the 2MASS images (both angles change
slowly with radius, but for simplicity we ignore these variations).
Although the near-infrared images clearly show that the central
20\arcmin\ of M33 have a larger inclination angle (58$\degr$) and a
smaller PA (14$\degr$), we use the kinematic values of these
parameters for our surface photometry to maintain consistency between
the photometry and the mass model.

\citet{c03} used a new CO map of M33 from the 14 m FCRAO telescope to
confirm the accuracy of the \hi\ data and further improve the angular
resolution in the inner part of the galaxy.  We use the CO data out to
a radius of 6~kpc and the \hi\ data at larger radii.  The combined CO
and \hi\ rotation curve has a resolution element that ranges from
45\arcsec\ to 3\farcm5 and features typical velocity uncertainties of
$\sim2$ \kms.  The overall mass distribution of M33 is very well
constrained by these data.

\begin{figure}[t!]
\epsscale{1.20}
\plotone{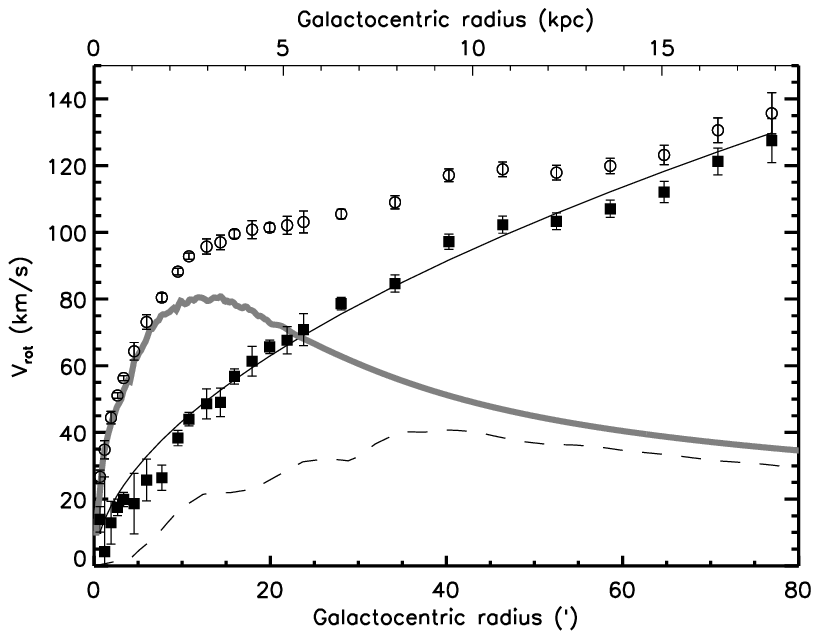}
\caption{Rotation curve of M33.  The open circles are the CO and \hi\
observations from \citet{c03} and \citet{cs00}.  The thick gray curve
represents the rotation curve of the stellar disk (for a stellar
mass-to-light ratio of $\mslk = 0.51 \mlk$), the dashed black curve
shows the rotation curve of the gas disk, and the black rectangles are
the rotation curve of the dark matter halo after removing the
contributions of the stars and gas.  The thin black solid curve is a
power law fit to the dark halo rotation curve.}
\label{rcfig}
\end{figure}

We construct a mass model of the galaxy assuming that it consists of a
spherical dark matter halo and thin disks of gas and stars.  The
stellar surface density profile is derived from the 2MASS $K_{s}$
image with the IRAF\footnote{IRAF is distributed by the National
  Optical Astronomy Observatories, which is operated by the
  Association of Universities for Research in Astronomy, Inc.  (AURA)
  under cooperative agreement with the National Science Foundation.}
task {\sc ellipse} in the STSDAS package, and the atomic and molecular
surface density profiles are taken from \citet{cs00} and \citet{c03},
respectively.  The rotation curves due to each of the baryonic
components are determined directly from the surface density profiles
by numerical integration using the NEMO software package
\citep{teuben}.  The observed CO and \hi\ surface density profiles
extend out to a radius of 80\arcmin\ (the edge of the rotation curve),
but the stellar surface densities can only be measured out to
20\arcmin.  To extend the rotation curve from the stellar disk out to
larger radii, we extrapolate the stellar surface densities using the
exponential disk fit from \S \ref{2mass}.  We assume a maximal stellar
disk, which for these data occurs at a stellar mass-to-light ratio of
$\mslk = 0.51 \mlk$.  It is worth noting that this value is in
agreement with the stellar M/L predicted from the galaxy color of
$\mslk = 0.55 \mlk$ \citep{bdj01}.  The rotational velocities due to
the dark halo are then calculated by subtracting in quadrature the
rotation curves of the stellar and gaseous disks from the observed
rotation curve (see Figure \ref{rcfig}).  To lessen the influence of
small-scale bumps and wiggles and ensure that the dark matter rotation
curve is monotonically increasing, we then replace the dark halo
rotation velocities with a power law fit.  We find that $v_{rot}
\propto r^{0.53}$ provides an excellent fit to the data.  Since the
dark halo is assumed to be spherical, the mass profile of the dark
matter is easily determined from its rotation curve.

To derive local values of the dynamical M/L, we divide each mass
component into annuli and add up the total mass contained in each
annulus.  We divide up the light profile in the same way, and then
take the ratio of the mass to the light in each annulus.  This M/L
profile can then be interpolated into a smooth function that gives the
M/L at any radius.  Note that unlike the surface brightness
measurements, these mass-to-light ratios are not truly local
measurements in the sense of being determined at a single position in
the galaxy.  Because of the nature of a rotation curve derived from a
two-dimensional velocity field, the M/L values are necessarily
azimuthally averaged quantities.

The uncertainties on the mass-to-light ratios that we derive deserve
some further comment here (see also \S \ref{relationships}).  Although
the uncertainties on the rotation curve and the light profile of the
galaxy are known, and therefore we can straightforwardly estimate the
uncertainty on the \emph{integrated} mass-to-light ratio interior to a
particular radius, the uncertainty on the \emph{local} mass-to-light
ratio at any given radius is less well defined since it is effectively
the ratio of differences between large numbers.  Because the other
galaxy properties that we are comparing with M/L are measured locally,
it is important that we use local mass-to-light ratios instead of
integrated ones, despite the potentially large uncertainties involved.

\section{RELATIONSHIPS BETWEEN M/L, METALLICITY, AND SURFACE BRIGHTNESS}
\label{relationships}

\begin{figure}[th!]
\epsscale{1.20}
\plotone{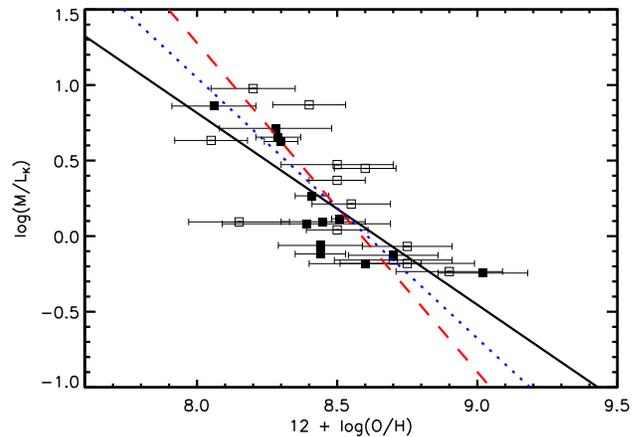}
\caption{M/L-metallicity relationship for \hii\ regions (filled
  squares) and supergiant stars (open squares) in M33.  The solid line
  represents the forward fit to the data, the dashed red line
  represents the fit to the inverse relationship, incorporating the
  metallicity uncertainties, and the blue dotted line is the linear
  bisector of the two (Equation \ref{ml_metallicityequation})}.
\label{ml_metallicityplot}
\end{figure}

\begin{figure}[th!]
\epsscale{1.20}
\plotone{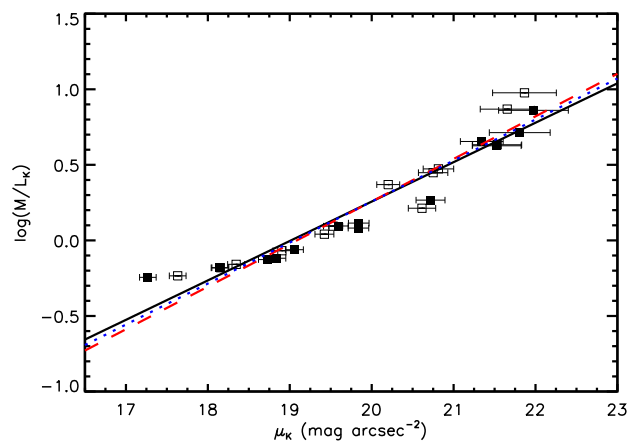}
\caption{M/L-surface brightness relationship for \hii\ regions (filled
  squares) and supergiant stars (open squares) in M33.  The solid line
  represents the fit to the data using M/L as the independent
  variable, the dashed red line represents the fit to the inverse
  relationship, incorporating the surface brightness uncertainties,
  and the blue dotted line is the linear bisector of the two (Equation
  \ref{ml_sbequation})}.
\label{ml_sbplot}
\end{figure}

\begin{figure}[th!]
\epsscale{1.20}
\plotone{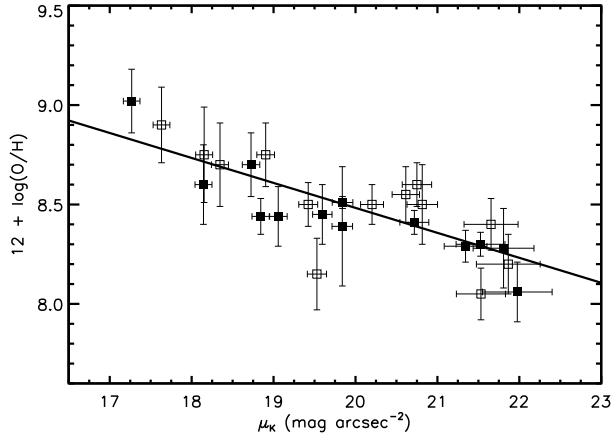}
\caption{Surface brightness-metallicity relationship for \hii\ regions
  (filled squares) and supergiant stars (open squares) in M33.  The
  solid line represents the fit to the data given in Equation
  \ref{metallicity_sb}.}
\label{sb_metallicityplot}
\end{figure}

In Figures \ref{ml_metallicityplot}, \ref{ml_sbplot}, and
\ref{sb_metallicityplot} we show that there are strong correlations
between all three variables under consideration.  We measure
correlation coefficients of $-0.74$ for the log M/L-metallicity
correlation, $0.96$ for the log M/L-surface brightness correlation,
and $-0.80$ for the metallicity-surface brightness correlation.  In
each case the trends go in the expected sense; metallicity and surface
brightness both decline with increasing M/L, and surface brightness
increases with metallicity.

Before we proceed to fitting the data, a brief discussion of some of
the subtleties involved is warranted.  We are in the somewhat unusual
situation of investigating relationships between three parameters, two
of which have reasonably well-known and well-understood uncertainties
(metallicity and surface brightness), and one of which does not
(mass-to-light ratio).  We do not know at this point whether one of
these parameters physically causes the correlations with the others,
so we should not assign the roles of independent and dependent
variables in the fits.  The most appropriate way of applying the
standard technique of linear least squares fitting is therefore not
clear.  As discussed by \citet{isobe90} and \citet{ab96}, there are a
number of fitting methods that can be used for linear regression
problems.  Because the relationships that we are investigating may
have intrinsic scatter that is as large as or larger than the
measurement uncertainties, we use the bivariate correlated errors and
intrinsic scatter (BCES) method proposed by \citet{ab96} as a
generalization of ordinary least squares fitting.  In order to avoid
biasing the results by designating one variable as independent and one
as dependent, for each pair of variables we run the BCES regression of
$Y$ on $X$ as well as the BCES regression of $X$ on $Y$, and then take
the linear bisector of the two as the best estimate of the true
relationship.

\subsection{Two-Parameter Correlations}
\label{correlations}

We begin our analysis by considering the two-parameter correlations
between M/L and metallicity, M/L and surface brightness, and
metallicity and surface brightness.  Following PB02, we fit the
mass-to-light ratios at the position of each metallicity measurement
as a function of the metallicities.  Using a BCES fit with M/L as the
dependent variable, we find the following relationship:
$log{(M/L_{K})} = 10.97 \pm 1.73 - (1.27 \pm 0.20)[12 + log(O/H)]$.
Because the metallicities have known uncertainties, while the
mass-to-light ratios do not, it might make more sense to carry out the
inverse fit with metallicity as the dependent variable, which yields
$12 + log(O/H) = 8.59 \pm 0.04 - (0.46 \pm 0.08)log(M/L_{K})$.
Converting this equation back into the form of the first equation, we
find $log{(M/L_{K})} = 18.69 \pm 3.17 - (2.18 \pm 0.37)[12 +
  log(O/H)]$, which differs from the forward fit at the
$\sim3$~$\sigma$ level in both coefficients.  The reduced $\chi^{2}$
for this fit is 1.35.  The forward and inverse fit results give a
sense of the range of fits that are compatible with the data.
Although there is a clear correlation between metallicity and
mass-to-light ratio, the large scatter makes it difficult to define a
precise relationship.  Because the inverse fit incorporates additional
information (the metallicity uncertainties), it is probably a more
reliable estimate of the true correlation.  The linear bisector (the
average of the first and third equations) of the metallicity-M/L
relationship is

\begin{equation}
log{(M/L_{K})} = 14.83 \pm 1.80 - (1.72 \pm 0.21)[12 + log(O/H)].
\label{ml_metallicityequation}
\end{equation}

\noindent The forward and inverse fits, along with the linear
  bisector, are plotted in Figure \ref{ml_metallicityplot}.

If we relax our assumption that the stellar disk of M33 is maximal and
construct mass models with lower stellar mass-to-light ratios, we find
only modest changes in the derived relationship.  Even for the
limiting case of a minimal disk (in which the stars do not contribute
any mass to the galaxy), the fit parameters change by at most $\sim
1.95~\sigma$, in the direction of increasing zero point and decreasing
(more negative) slope.

We then carry out the same process with the surface brightnesses.  For
the unweighted fit with M/L as the dependent variable, we find
$log{(M/L_{K})} = -4.96 \pm 0.34 + (0.26 \pm 0.02)\mu_{K}$.  The
inverse fit (including surface brightness uncertainties), gives
$\mu_{K} = 19.09 \pm 0.10 + (3.55 \pm 0.23)log{(M/L_{K})}$ and when we
rearrange this equation to write it in terms of M/L, we get
$log{(M/L_{K})} = -5.38 \pm 0.34 + (0.28 \pm 0.02)\mu_{K}$, this time
in reasonably good agreement with the forward relationship.  The
formal reduced $\chi^{2}$ value for this fit is quite high (9.3)
because of the small errors on the surface photometry near the center
of the galaxy, indicating that the observed scatter is intrinsic to
the relationship.  Nevertheless, it is clear that the fitted lines are
a good description of the data (see Figure \ref{ml_sbplot}).  The
linear bisector of the surface brightness-M/L correlation is

\begin{equation}
log{(M/L_{K})} = -5.17 \pm 0.24  + (0.27 \pm 0.01)\mu_{K}.
\label{ml_sbequation}
\end{equation}

For the M/L-surface brightness relationship, making the stellar disk
less massive does have a statistically significant effect on the
derived parameters.  The linear bisector correlation for a submaximal
($V_{disk}/V_{tot} = 0.6$) disk with $\mslk = 0.27 \mlk$ is
$log{(M/L_{K})} = -5.76 + 0.30\mu_{K}$, and for a minimal stellar disk
the correlation is $log{(M/L_{K})} = -6.50 + 0.34\mu_{K}$.

Finally, for the metallicity-surface brightness relation, we obtain a
best fit (taking uncertainties in both parameters into account) of
\begin{equation}
12 + log(O/H) = 10.99 \pm 0.40 - (0.13 \pm 0.02)\mu_{K},
\label{metallicity_sb}
\end{equation}

\noindent
with a reduced $\chi^{2}$ value of 1.02 (see Figure
\ref{sb_metallicityplot}).  Note that in this case, since the observed
scatter is consistent with being entirely attributable to the
observational uncertainties, we do not need to use the BCES technique.
We instead employ the method of \citet{nr} for linear fitting with
errors in both coordinates.  The inverse fit gives identical results
for this relationship (which is not surprising, since the variables
are treated completely symmetrically).  This result is in excellent
agreement with the $I$-band relationship derived by \citet{ryder95}
from measurements of 97 \hii\ regions in six galaxies,
\begin{equation}
12 + log(O/H) = 10.91 \pm 0.90 - (0.095 \pm 0.034)\mu_{I}.
\label{metallicity_sb_ryder}
\end{equation}

\subsection{The Fundamental Line}
\label{fl}

Having established the various two-parameter correlations between the
data points, we now consider the best-fitting plane for all three
parameters simultaneously.  Using an unweighted ordinary least squares
fit, we find that the fundamental line can be expressed as
\begin{eqnarray}
\nonumber log{(M/L_{K})} & = & -6.62 \pm 1.42 \\
 \label{flfit}
                        &   & + (0.15 \pm 0.16)[12 + log(O/H)] \\
\nonumber                &   & + (0.28 \pm 0.03)\mu_{K}.
\end{eqnarray}

\begin{figure}[t!]
\epsscale{1.20}
\plotone{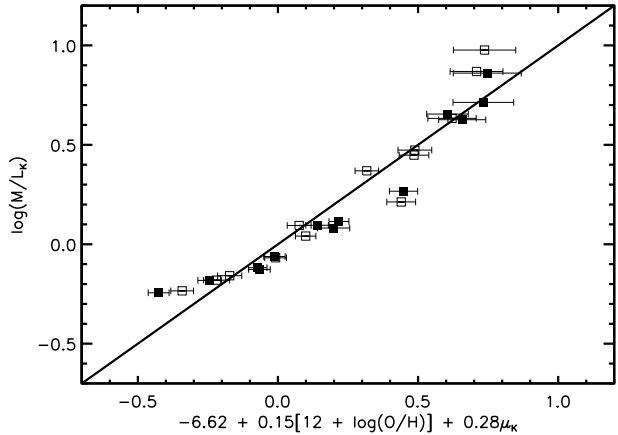}
\caption{The local fundamental line within M33.  The filled squares
  represent \hii\ regions and the open squares represent 
  supergiants.  The solid line shows the fit to the data given in
  Equation \ref{flfit}.}
\label{flplot}
\end{figure}

\begin{figure*}[th!]
\epsscale{1.17}
\plottwo{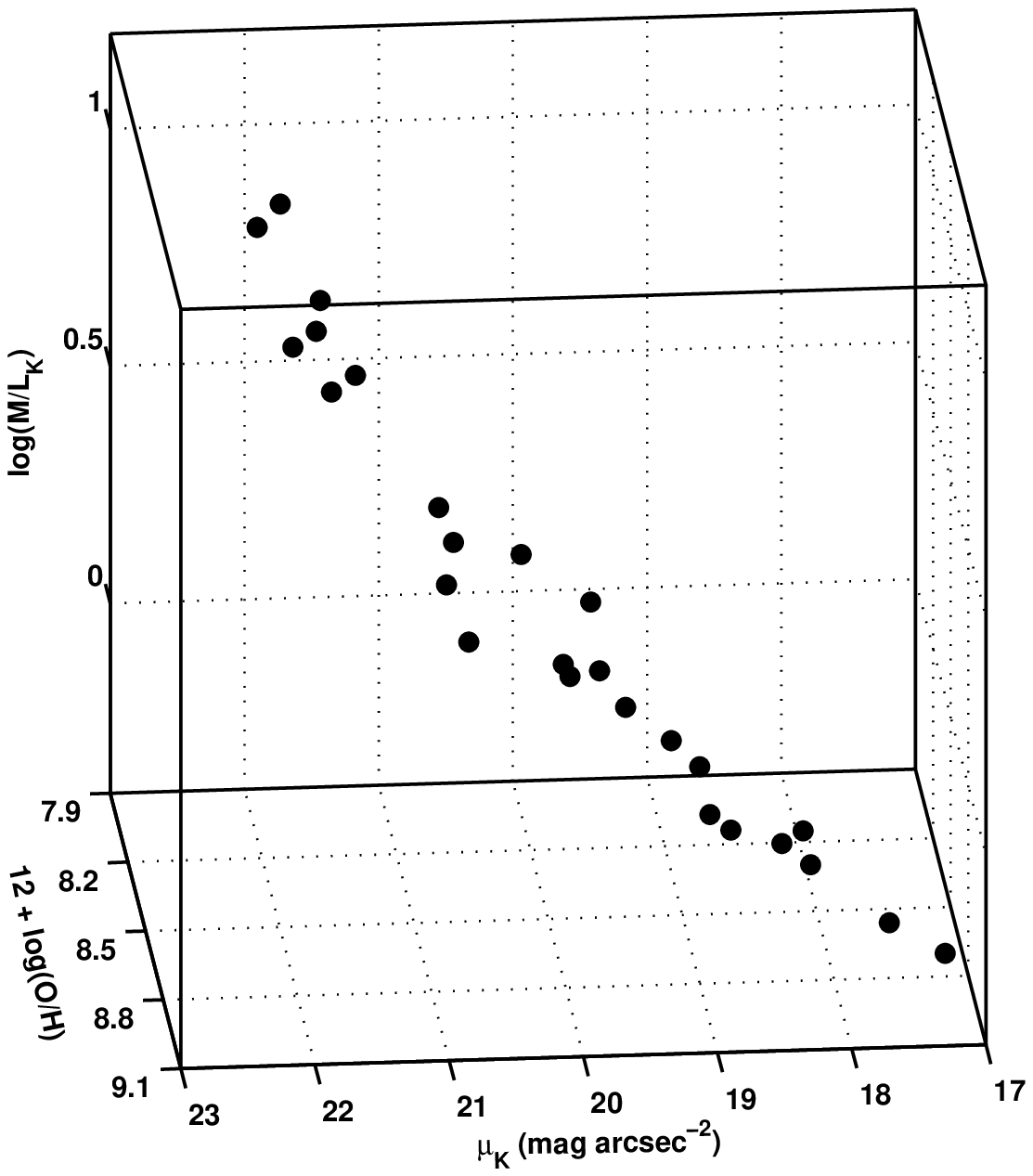}{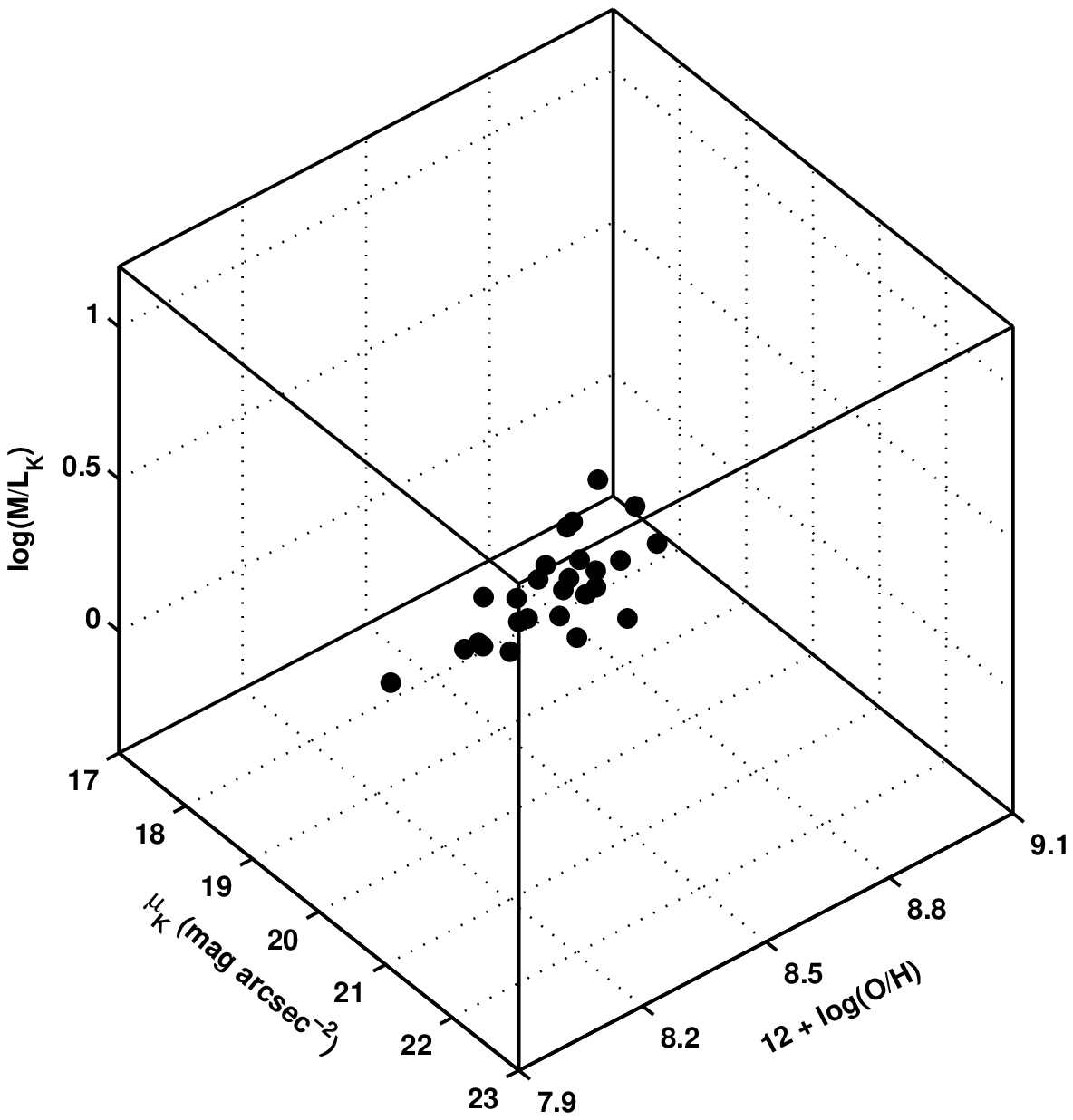}
\caption{Edge-on (left) and end-on (right) views of the fundamental
  line in the three-dimensional space (metallicity, surface
  brightness, log(M/L)).  The narrowness of the fundamental line when
  observed end-on demonstrates that the points are indeed distributed
  in a line rather than a plane.}
\label{three-d} 
\end{figure*}

\noindent
The local fundamental line for M33 is plotted in Figure
\ref{flplot}. The effect of increasing the assumed distance for M33
from 800~kpc to 900~kpc (see \S \ref{abundances}) is to slightly
increase the constant offset in Equation \ref{flfit} to $-6.47$.  The
metallicity coefficient is unchanged, and the surface brightness
coefficient decreases to 0.27; all three terms change by much less
than $1~\sigma$.  Adjusting the mass model to allow for a lower mass
stellar disk also does not significantly change the fundamental line.
Even for a minimal stellar disk the constant term and the metallicity
coefficient change by less than their uncertainties, and the surface
brightness coefficient increases by $1.7~\sigma$.  The derived
fundamental line therefore appears to be robust against plausible
systematic uncertainties in our assumptions.  In Figure \ref{three-d}
we display edge-on and end-on views of the fundamental line in three
dimensions, proving that the relationship is indeed a line and not a
plane.

In contrast with the dwarf galaxy data, we find a very weak
metallicity dependence for the fundamental line in M33.  The
metallicity dependence is also in the opposite sense of what we found
when considering the two-parameter M/L-metallicity relationship.  This
may be partially caused by the small range of metallicities included
in our sample ($\sim1$ order of magnitude) and the relatively large
uncertainties on those metallicities, but could also indicate a real
difference between M33 and the Local Group dwarfs.  In support of the
former interpretation, it is worth noting that the M/L-metallicity
correlation is the weakest of the three relationships studied in \S
\ref{correlations}.  Measurements of a large sample of metallicities
including the innermost and outermost regions of M33 may be able to
determine whether the metallicity dependence of the disk galaxy and
dwarf galaxy fundamental lines actually differ.

\subsubsection{The Curvature of the Fundamental Line}
\label{curvature}

Despite the name that we have chosen for the fundamental line,
inspection of Figure \ref{flplot} suggests that rather than being
truly linear, the fundamental line is actually slightly curved.  At
both the low M/L and high M/L ends of the line the data are curving
upwards, towards larger mass-to-light ratios than the linear relation
would predict.  These deviations from linearity can be naturally
understood by considering what the local values of M/L are physically
describing.  Near the center of the galaxy (at low M/L), the baryons
contribute substantially to the total mass, increasing the
mass-to-light ratio.  If we used only the dark matter halo to
calculate masses, then this end of the line would be more nearly
straight (as we show explicitly in \S \ref{origin}).  The other end of
the fundamental line also curves upward because there is almost no
light being added at large radii, making the mass-to-light ratio
increase more rapidly.  Consistent with these explanations, we note
that similar curvature is evident in the M/L-metallicity and
M/L-surface brightness correlations, but not in the
metallicity-surface brightness relationship, indicating that the
origin of the curvature is in the mass-to-light ratio.  Because of the
limited extent of the available data, it is more useful to use the
linear description of the fundamental line that we have derived rather
than a quadratic one.  However, future studies that employ larger data
sets may be able to examine the curvature of the relationship in more
detail.

\section{DISCUSSION}
\label{discussion}

In the previous section, we showed that not only are mass-to-light
ratio, metallicity, and surface brightness all correlated with each
other in M33, but also that a local fundamental line, which holds at
every measured point within the disk of the galaxy, exists as well.
We now examine this relationship in comparison to the dwarf galaxy
fundamental line and consider its implications.

\subsection{Comparison with the Fundamental Line of Dwarf Galaxies}
\label{dwarfs}

Since the motivation for this work was the discovery of the
fundamental line of dwarf galaxies, we would like to compare the
relationships that we have found within M33 with the analogous
relationships among dwarfs.  Using linear least squares fits to the
forward relationships, PB02 derived the following three correlations
for the integrated properties of Local Group dwarfs:
\begin{equation}
log{(M/L_{V})} = -0.4 - [Fe/H],
\label{ml_metallicity_dwarfs}
\end{equation}
\begin{equation}
log{(M/L_{V})} = -5.35 + 0.27\mu_{V},
\label{ml_sb_dwarfs}
\end{equation}

\noindent and
\begin{equation}
log{(M/L_{V})} = -2.86 - 0.56[Fe/H] + 0.13\mu_{V}.
\label{fl_dwarfs}
\end{equation}

If we fit the same data using the BCES method, as we did for M33 in \S
\ref{correlations}, we find that the coefficients differ at the
$\sim1-2$~$\sigma$ level:
\begin{equation}
log{(M/L_{V})} = -0.68 \pm 0.13 - (1.13 \pm 0.09)[Fe/H]
\label{ml_metallicity_dwarfs_bces}
\end{equation}

\noindent and
\begin{equation}
log{(M/L_{V})} = -6.23 \pm 0.53 + (0.31 \pm 0.02)\mu_{V}
\label{ml_sb_dwarfs_bces}
\end{equation}

In order to compare these fits properly with our results, we must
derive conversions between the [Fe/H] and [12 + log(O/H)] abundance
scales and $V$- and $K$-band mass-to-light ratios and surface
brightnesses.  A subset of the Local Group dwarf galaxies discussed by
\citet{mateo98} have both stellar [Fe/H] measurements and \hii\ region
oxygen abundances, from which we obtain
\begin{equation}
12 + log(O/H) = 9.5 \pm 0.5 + (1.1 \pm 0.4)[Fe/H],
\label{fe_to_o1}
\end{equation}

\noindent
or alternatively,
\begin{equation}
[Fe/H] = -8.5 \pm 2.6 + (0.9 \pm 0.3)[12 + log(O/H)].
\label{fe_to_o2}
\end{equation}

\begin{figure*}[th!]
\epsscale{1.17}
\plottwo{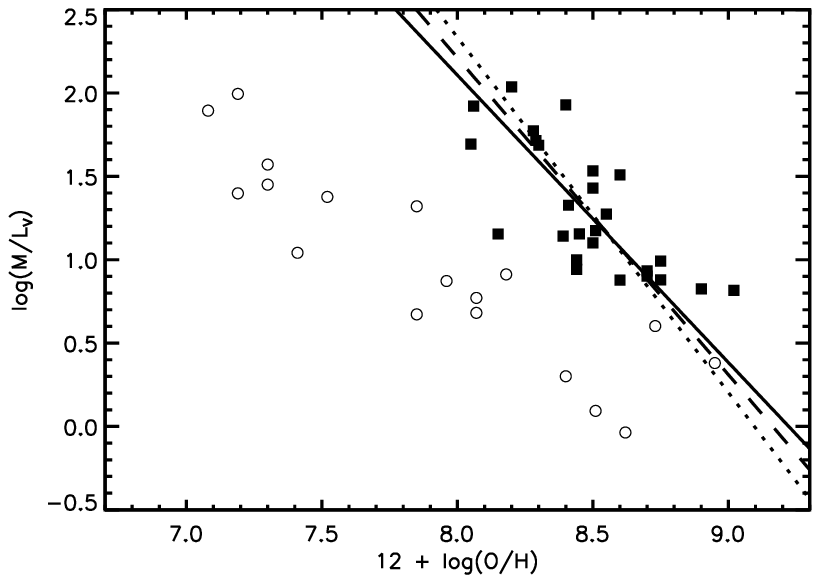}{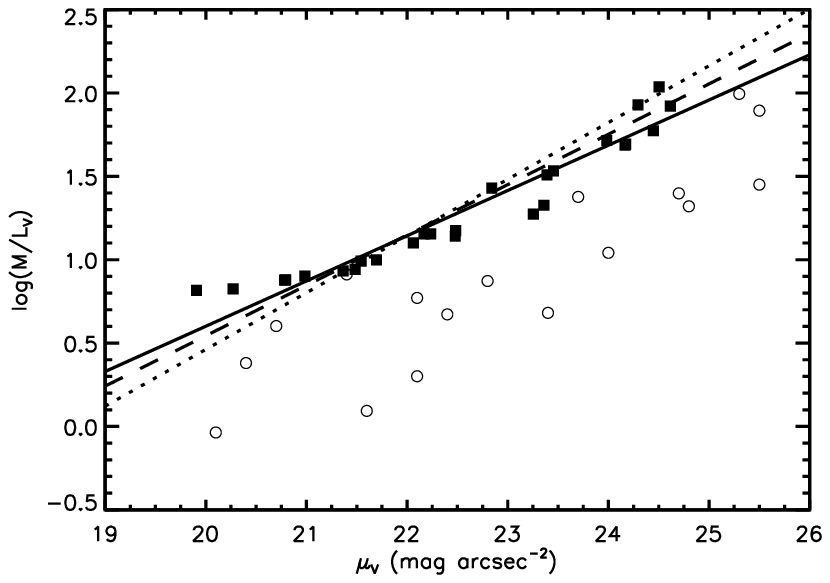}
\caption{Comparison of the M/L-metallicity (left panel) and
  M/L-surface brightness (right panel) relationships within M33 and
  among the Local Group dwarfs.  The M33 data are plotted as filled
  squares, and the dwarf galaxy points are plotted as open circles.
  The solid lines show the linear bisector fits to the M/L-metallicity
  and M/L-surface brightness relationships from Figures
  \ref{ml_metallicityplot} and \ref{ml_sbplot}.  The dashed
  (submaximal disk with $\mslk = 0.27$~\mlk) and dotted (minimal disk
  with $\mslk = 0$~\mlk) lines illustrate the effect of using
  different mass models for M33.  Lowering the assumed stellar
  mass-to-light ratio does not improve the agreement between the M33
  and dwarf galaxy data points.}
\label{comp1}
\end{figure*}

\noindent
The abundance range spanned by the Local Group dwarf galaxies
encompasses the abundance range of our data in M33 (see Figure
\ref{comp1}).

$K$-band observations do not exist for many of the Local Group dwarfs,
so the $V$- to $K$-band conversions must be derived specifically from
the M33 data.  M33 has integrated magnitudes of $m_{V} = 5.37$
\citep{rc3} and $m_{K} = 2.73$ (both measurements are corrected for
Galactic and internal extinction according to \citeauthor{sfd98}
[1998] and \citeauthor{sakai00} [2000], respectively), giving a $V -
K_{s}$ color of 2.64.  We can therefore estimate that

\begin{equation}
\log{M/L_{V}} = \log{M/L_{K}} + 1.06
\label{v_to_k1}
\end{equation}

\noindent
and
\begin{equation}
\mu_{V} = \mu_{K} + 2.64.
\label{v_to_k2}
\end{equation}

Applying the conversion given in Equation \ref{fe_to_o2} to the dwarf
galaxy metallicities, we find that the dwarf galaxy M/L-metallicity
relation (Equation \ref{ml_metallicity_dwarfs_bces}) can be written

\begin{equation}
log{(M/L_{V})} = 8.93 \pm 3.04 - (1.02 \pm 0.35)[12 + log(O/H)].
\label{ml_metallicity_dwarfs2}
\end{equation}

\noindent
Converting the M33 relationships to $V$-band using Equations
\ref{v_to_k1} and \ref{v_to_k2} gives M/L-metallicity and M/L-surface
brightness relations of

\begin{equation}
log{(M/L_{V})} = 15.89 \pm 1.80 - (1.72 \pm 0.21)[12 + log(O/H)]
\label{ml_metallicity_m332}
\end{equation}

\noindent
and

\begin{equation}
log{(M/L_{V})} = -4.82 \pm 0.24 + (0.27 \pm 0.01)\mu_{V}.
\label{ml_sb_m332}
\end{equation}

\begin{figure}[t!]
\epsscale{1.20} \plotone{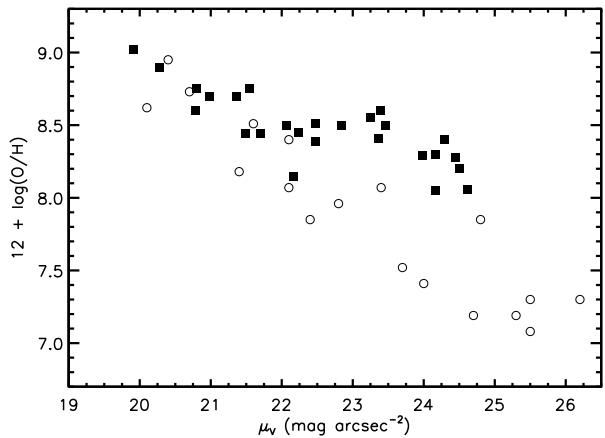}
\caption{Comparison of the surface brightness-metallicity
relationships within M33 and among the Local Group dwarfs.  The M33
data are plotted as filled squares, and the dwarf galaxy points are
plotted as open circles.
\label{comp3}}
\end{figure}

\noindent
Comparing Equations \ref{ml_metallicity_dwarfs2} and
\ref{ml_metallicity_m332}, we see that both the slope and the zero
point of the M/L-metallicity relationship are only marginally
consistent (at the $\sim2$~$\sigma$ level) between the Local Group
dwarf galaxies and M33.  The agreement between the global and local
M/L-surface brightness relationships (Equations
\ref{ml_sb_dwarfs_bces} and \ref{ml_sb_m332}) is slightly worse, with
a slope difference of $2$~$\sigma$ and a zero point offset of
$2.7$~$\sigma$.  However, as shown in Figures \ref{comp1} and
\ref{comp3}, for all three correlations the trends in the M33 data do
appear roughly consistent with the trends among the Local Group dwarfs
except for a zero-point offset.  The local M/L in M33 is roughly a
factor of 3 higher than the global M/L of a dwarf galaxy at the same
surface brightness, and a factor of $5-10$ higher at the same
metallicity.  It is worth noting that using different mass models for
M33 that place less mass in the stellar disk does not improve the
agreement between the local and global relationships.

\begin{figure}[t!]
\epsscale{1.20}
\plotone{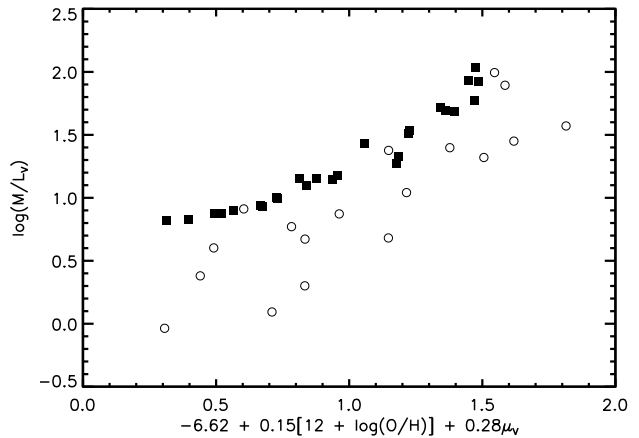}
\caption{Comparison of the fundamental line within M33 and the
  fundamental line of the Local Group dwarfs in the same units.  The
  M33 data are plotted as filled squares, and the dwarf galaxy points
  are plotted as open circles.
\label{flcomparison}}
\end{figure}

Converting the M33 and dwarf galaxy fundamental lines to the common
system of oxygen abundances and $V$-band magnitudes yields

\begin{equation}
log{(M/L_{V})} = 1.90 - 0.50[12 + \log{(O/H)}] + 0.13\mu_{V}
\label{fldwarf_m33units}
\end{equation}

\noindent
for the dwarf galaxies, and 

\begin{equation}
log{(M/L_{V})} = -6.30 \pm 1.42 + (0.15 \pm 0.16)[12 + \log{(O/H)}] +
(0.28 \pm 0.03)\mu_{V}
\label{flm33_dwarfunits}
\end{equation}

\noindent
for M33.  These relationships are clearly in rather poor agreement
with each other.  Nevertheless, a visual comparison again demonstrates
that the variation of the mass-to-light ratio with metallicity and
surface brightness is similar in M33 as it is for the dwarf galaxies
(see Figure \ref{flcomparison}).

The similarities in the slopes of all of these relationships indicate
that there could be a connection between the local and global
fundamental lines.  However, M33 does not lie on the fundamental line
of dwarfs, so the nature of this connection is not clear.  Because
structure forms hierarchically in the currently-favored Cold Dark
Matter (CDM) cosmology, disk galaxies such as M33 were presumably
constructed from the merging of many smaller galaxies.  We suggest
that simple simulations could be used to test whether a galaxy that is
built up of smaller objects that obeyed the fundamental line of dwarf
galaxies follows a local fundamental line like the one we observe in
M33 at the present day.  If not, what additional information is
necessary to create the fundamental line?

\subsection{The Origin of the Fundamental Line Correlations}
\label{origin}

\subsubsection{The Metallicity-Surface Brightness Relation}

The correlation between local surface brightness and metallicity in
disk galaxies is well-known from previous work
\citep[e.g.,][]{ep84,ryder95,garnett97,bdj00}.  This relationship is
essentially a generic result of ordinary chemical evolution models.
For example, if we treat M33 as a closed box, then the heavy element
fraction $Z$ as a function of the gas fraction $\mu$ can be written as

\begin{equation}
Z(\mu) = -p\ln\mu,
\label{closedboxev1}
\end{equation}

\noindent
where $p$ is the yield of heavy elements from each generation of star
formation \citep{ss72}.  If we write out the expressions for $Z$ and
$\mu$, then we have

\begin{equation}
\frac{M_{Z}}{M_{gas}} = -p\ln\left(\frac{M_{gas}}{M_{gas} + M_{*}} \right),
\label{closedboxev2}
\end{equation}

\noindent
where $M_{Z}$ is the total mass of heavy elements and $M_{*}$ is the
stellar mass.  Now, we divide the mass terms on the right-hand side of
the equation by a unit area to turn them into surface densities:

\begin{equation}
\frac{M_{Z}}{M_{gas}} = -p\ln\left(\frac{\Sigma_{gas}}{\Sigma_{gas} + \Sigma_{*}} \right).
\label{closedboxev3}
\end{equation}

\noindent
After substituting the oxygen mass for the total heavy element mass,
using $M_{O} = 0.45 M_{Z}$ \citep{garnett97}, and noting that $M_{gas}
= 1.33 M_{H}$, we can rearrange Equation \ref{closedboxev3} to obtain
the following for the metallicity-surface brightness relationship:

\begin{equation}
12 + \log{(O/H)} = \log{\left[p\ln{\left(1 + \frac{\Sigma_{*}}{\Sigma_{gas}}
      \right)} \right] + 11.78}.
\label{closedboxev4}
\end{equation}

\noindent
Solving this equation for the yield $p$ gives us

\begin{equation}
p = \frac{10^{12 + \log{O/H} - 11.78}}{\ln{\left( 1 + \Sigma_{*}/\Sigma_{gas}\right)}},
\label{closedboxev5}
\end{equation}

\noindent
and, using the data from \S \ref{relationships}, we find that for $p =
3.5 \times 10^{-4}$ this model reproduces the observed
metallicity-surface brightness correlation.

Alternatively, we can abandon the closed-box assumption and allow the
infall of metal-poor gas onto M33.  In this case, \citet{clayton87}
has shown that the heavy element fraction is

\begin{equation}
Z(\mu) = \frac{p}{2}\left[-\ln\mu + \ln\left(-\ln\mu + 1\right)\right].
\label{openboxev1}
\end{equation}

\noindent
Proceeding through the same steps as above, we find that 

\begin{equation}
p = \frac{10^{12 + \log{O/H} - 11.48}}
{ -\ln{\left( \frac{\Sigma_{gas}}{\Sigma_{gas} + \Sigma_{*}}\right)}
 + \ln{\left[ 1 - \ln{ \left(\frac{\Sigma_{gas}}{\Sigma_{gas} + \Sigma_{*}}\right)}\right]} }.
\label{openboxev2}
\end{equation}

\noindent
In this case also, for a yield of $p = 4.5 \times 10^{-4}$, we
correctly recover the surface brightness-metallicity relationship in
M33.  It appears likely, then, that most plausible chemical enrichment
scenarios will produce a strong correlation between metallicity and
surface brightness.  We therefore expect that all disk galaxies should
display such a correlation.  Galaxies that lie on the same
correlation, as M33 and the sample studied by \citet{ryder95} do, may
have similar values for the heavy element yield.

\subsubsection{The M/L-Surface Brightness Relation}

Relationships between surface brightness and mass-to-light ratio have
been recognized before on a global level \citep[e.g.,][although see
  \citealt{graham02}]{zwaan95,zavala03}, but the corresponding
relationship within a single galaxy has received only minimal
attention \citep{petrou81}.  In order to understand the origin of the
local M/L-surface brightness correlation that we found in \S
\ref{correlations}, we consider a galaxy with an exponential stellar
disk and a \citet[][hereafter NFW]{nfw96} dark matter halo:

\begin{eqnarray}
I(r) & = & I_{0}e^{-r/r_{d}} \\
\rho(r) & = & \frac{\delta_{c}\rho_{c}}{\frac{r}{r_{s}}\left(1 + \frac{r}{r_{s}}\right)^{2}}.
\end{eqnarray}

\noindent
Integrating the surface brightness profile over the circular disk area and the
density profile over a spherical volume give the integrated luminosity
and mass profiles of the galaxy, respectively:

\begin{eqnarray}
L(r) & = & 2\pi I_{0}r_{d}^{2}\left[ 1 - e^{-r/r_{d}} \left( 1 + 
           \frac{r}{r_{d}}\right) \right] \label{lr}\\
M(r) & = & 4\pi\delta_{c}\rho_{c}r_{s}^{3}\left[ \ln{\left(1 +
           \frac{r}{r_{s}}\right)} - \frac{\frac{r}{r_{s}}}{1 + 
           \frac{r}{r_{s}}} \right]. 
\end{eqnarray}

\begin{figure}[th!]
\epsscale{1.20}
\plotone{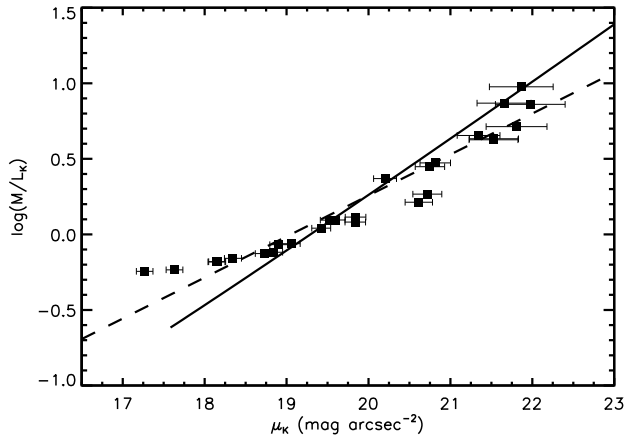}
\caption{Model M/L-surface brightness relationship for a galaxy with
  an NFW dark matter halo and an exponential disk.  The solid line
  shows the log of Equation \ref{localml} plotted against surface
  brightness, using parameters for the dark matter halo and stellar
  disk that are appropriate for M33.  The solid points show the
  observed data for M33, and the dashed line gives the M/L-surface
  brightness relationship that we derived from the data in Equation
  \ref{ml_sbequation}.}
\label{ml_sbmodel}
\end{figure}

Dividing the galaxy into concentric rings of width $\Delta r$, calculating
the incremental mass ($\Delta M = M(r+\Delta r) - M(r)$) and light
($\Delta L = L(r+\Delta r) - L(r)$) in each ring using Equations
\ref{lr} and 28, and then taking the ratio of the two yields

\begin{equation}
\frac{\Delta M}{\Delta L} =
\frac{2\delta_{c}\rho_{c}r_{s}^{3}}{I_{0}r_{d}^{2} e^{-r/r_{d}}}
\frac{\ln{(1 + \frac{r + \Delta r}{r_{s}})} - \ln{(1 + \frac{r}{r_{s}})}
      + \frac{r}{r + r_{s}} - \frac{r + \Delta r}{r + \Delta r + r_{s}}}
{1 + \frac{r}{r_{d}} - e^{-\Delta r/r_{d}}\left( 1 + \frac{r + \Delta r}{r_{d}}\right) }
\label{localml}
\end{equation}

\noindent
for the local mass-to-light ratio.  While this expression may look
formidable, in fact its logarithm scales linearly with surface
brightness (and radius) as long as $r_{s} \gg r_{d}$ (which is always
true for real galaxies).  Using reasonable values for the disk of M33
from this work and the halo from \citet{c03} (NFW concentration $c =
10$ and scale radius $r_{s} = 15$~kpc), we plot the M/L-surface
brightness relation according to Equation \ref{localml} in Figure
\ref{ml_sbmodel}.  There is reasonable agreement between the modeled
and observed relationships.  If the model had included the baryonic
contribution to the total mass, which is significant at small radii
(high surface brightnesses), then the solid line would curve upward at
$\mu_{K} \lesssim 19$, following the data.  Using other density
profiles for the dark matter halo, such as a power law (as we assumed
in \S \ref{rc}) or a pseudo-isothermal profile, does not change the
essentially linear correlation between $\log{M/L}$ and surface
brightness, although in both of those cases the relationship does bend
slightly upwards at high surface brightnesses rather than remaining
exactly linear.  Thus, the M/L-surface brightness correlation appears
to be a universal property of galaxy disks that simply reflects the
exponential nature of the disk and the increasing dark matter fraction
as a function of radius.

\subsubsection{The M/L-Metallicity Relation}

Although the metallicity-surface brightness and M/L-surface brightness
relationships follow generically from the basic processes that occur
in galaxies, what is not clear is whether the M/L-metallicity
relationship likewise is fundamental to the nature of disk galaxies,
or if it is merely a residual correlation resulting from the other two
relationships.  Since the M/L-metallicity relationship has the largest
scatter of the three, one might be justified in suspecting the latter.
However, as we discussed in \S \ref{introduction}, PB02 proposed a
model that can reproduce the global M/L-metallicity relationship in
dwarfs, and simulations confirm that this result is plausible
\citep{tassis03}.  Therefore, it may be the case that there is a real
local M/L-metallicity relationship in disks, but the relationship is
weakened by the relatively large uncertainties on existing metallicity
measurements.  Future metallicity studies should be able to shed light
on this possibility.

\subsection{Implications}
\label{implications}

Previous work suggested that feedback might be responsible for the
fundamental line of dwarf galaxies.  \citet{dw03} studied the scaling
relations among stellar mass, surface brightness, characteristic
velocity, and metallicity for both the Local Group dwarfs and
low-luminosity galaxies in the Sloan Digital Sky Survey database.
They argued that a simple recipe for supernova feedback could
reproduce the dwarf galaxy fundamental line.  \citet{tassis03} used
cosmological hydrodynamic simulations to show that a similar feedback
model produces a M/L-metallicity relationship for dwarf galaxies
consistent with that found by PB02.  It therefore seems plausible that
the same type of model might be able to explain the local fundamental
line within M33 (note that M33 does lie within the mass range for
which the \citeauthor{dw03} scaling relations apply, with a stellar
mass of $4 \times 10^{9}$~M$_{\odot}$, compared to the mass of $3
\times 10^{10}$~M$_{\odot}$ where the relationships change form).
However, our analysis in \S \ref{origin} indicates that at least two
of the three correlations that comprise the local fundamental line can
be derived \emph{without} appealing to feedback.  Furthermore, other
numerical work shows that both the local and global fundamental lines
can be produced in simulations that do not contain feedback (\S
\ref{simulations} of this paper and A. Kravtsov 2005, personal
communication).

If the \citet{dw03} picture is correct, then only galaxies below
the transition mass of $\sim3 \times 10^{10}$~M$_{\odot}$ in stars
($\sim6 \times 10^{11}$~M$_{\odot}$ in total mass) should exhibit a
local fundamental line.  For more massive systems, \citet{dekel04}
suggests that feedback from active galactic nuclei plays an important
role in shaping the evolution of galaxies.  A critical test of the
supernova feedback model is therefore whether a fundamental line
relationship holds for massive, early-type disk galaxies and massive
ellipticals.  This test could be carried out for nearby galaxies with
photometry of large samples of red giant branch stars (to measure the
mean metallicity as a function of radius) along with long-slit
spectroscopy to obtain the velocity dispersion profile (and hence
M/L).

A further implication of the local fundamental line is that if low
surface brightness (LSB) disk galaxies follow the fundamental line,
then they should have large mass-to-light ratios and low
metallicities.  Observations clearly indicate that LSB galaxies are
metal-poor \citep[e.g.,][]{dn04}, and there is also evidence for higher
mass-to-light ratios than are found in normal galaxies
\citep[][although see \citealt{sprayberry95}]{zwaan95,dbm96}

Finally, the fundamental line of M33 can also be used to place
constraints on CDM simulations of the formation of disk galaxies.
Hydrodynamic simulations of galaxy disks are beginning to approach the
actual properties of observed disks in terms of size, mass, and
rotation velocity, although the formation of bulgeless disks such as
M33 is rare \citep[e.g.,][]{abadi03,robertson04,governato04}.  The
results of this paper show that realistic simulations of disks must
also obey the fundamental line; if they do not, then the star
formation and/or feedback recipes they include must not be correct.
Current simulations are already detailed enough to investigate
mass-to-light ratios, surface brightnesses, and metallicities on small
physical scales, so in the following subsection we carry out this
test.

\subsection{The Fundamental Line in Simulations}
\label{simulations}

We analyzed the fundamental line properties of the $\Lambda$CDM disk
galaxy simulation of \citet{robertson04}, updated using the GADGET2
code \citep{gadget2}.  The simulation uses the \citet{sh03} model for
the multiphase ISM to explore the impact of ISM pressurization by
supernova feedback on disk galaxy formation, and produces a disk
galaxy with a mass of $2.16 \times 10^{11} h^{-1}$~M$_{\odot}$ and an
exponential surface brightness profile (see \citealt{robertson04} for
further details).

We extracted mass-to-light ratios, K-band surface brightnesses, and
gas metallicities from the simulation as follows.  We calculated the
angular momentum of the disk from the gas in order to define a face-on
disk plane.  Our measurements were then made in the plane of the disk.
The K-band luminosity of the stellar particles was determined from the
\citet{bc03} population synthesis models.  The metallicity and surface
brightness were measured in azimuthally averaged annuli, so these
quantities are not truly local values as our observations are.
However, the simulated galaxy is fairly axisymmetric (apart from a
weak bar feature), so we do not expect this difference to affect the
results.  The mass-to-light ratios were calculated by directly
measuring the dynamical mass in spherical shells and comparing with
the luminosity contained in annuli at the same radii, essentially the
same method that we used for M33.

\begin{figure*}[th!]
\epsscale{1.20}
\plotone{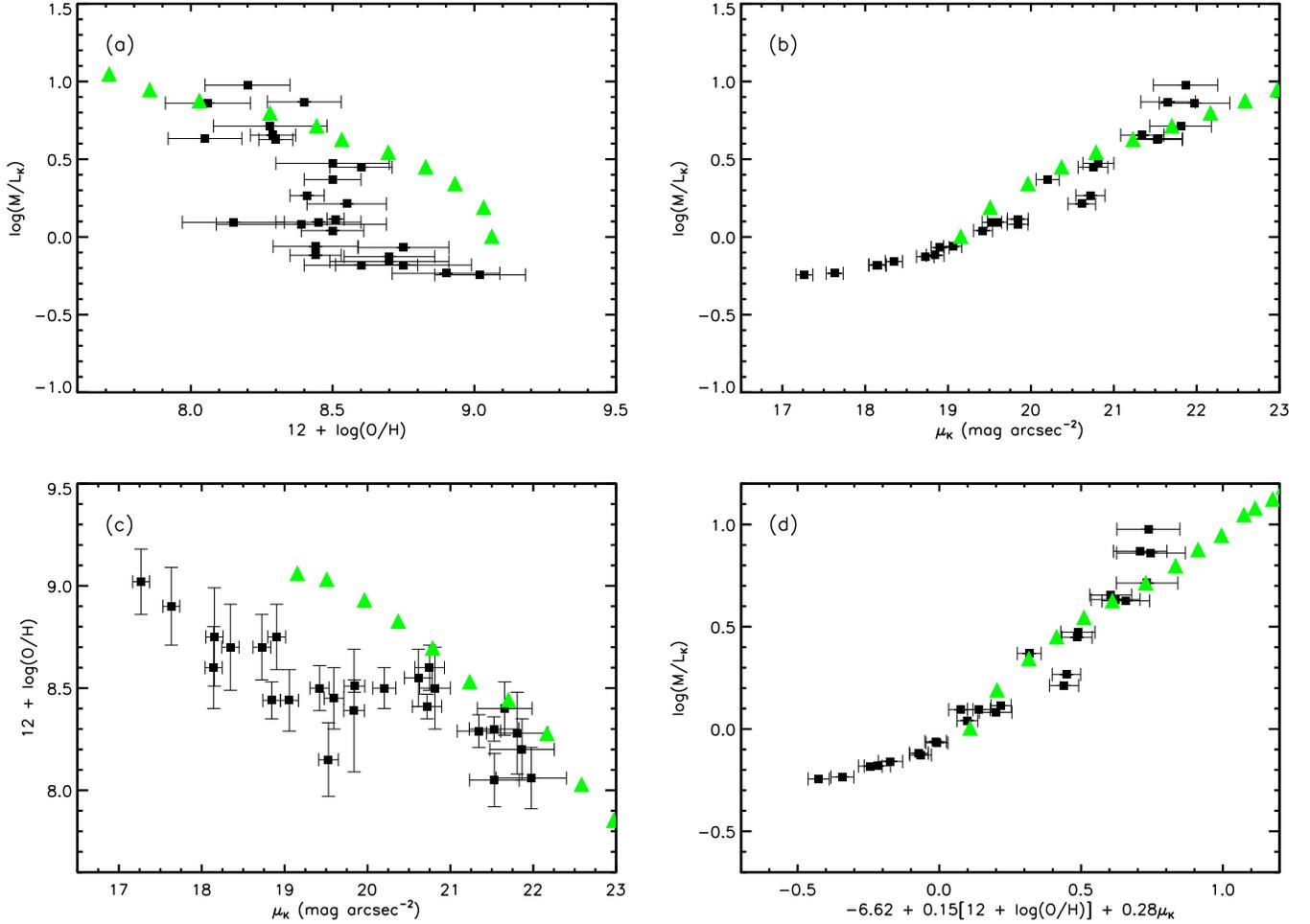}
\caption{Comparison of the scaling relationships in a simulated galaxy
  from \citet{robertson04} with the observed relationships in M33.
  \emph{(a)} The M/L-metallicity relationship.  \emph{(b)} The
  M/L-surface brightness relationship.  \emph{(c)} The surface
  brightness-metallicity relationship.  \emph{(d)} The fundamental
  line.  The black squares are the M33 data points, and the green
  triangles are the measurements of the simulated galaxy.  All of the
  trends in the data are correctly reproduced in the simulation,
  although there is a significant zero-point offset in the simulation
  metallicities.
\label{simulation}}
\end{figure*}

We plot the M/L-metallicity, M/L-surface brightness,
metallicity-surface brightness, and fundamental line relationships for
the simulated galaxy in Figure \ref{simulation}.  All of the trends we
found in the M33 data in \S \ref{relationships} are matched by the
simulation, although there are also clear offsets between some of the
actual relationships and the simulated ones, most notably involving
the metallicity.  Remarkably, though, the fundamental line is
reproduced very nearly correctly in the simulation.

The gas metallicities seen in the simulation are too large by
$\sim0.6$~dex, which is not surprising because the simulation does not
include a model for strong supernova-driven gas outflows.  Also,
simulated low-mass galaxies typically form too many stars given their
total mass and gas content, thereby producing too many metals as
well.\footnote{The reason that low-mass galaxies form too many stars
  in simulations may be that the star formation prescriptions are
  extrapolated from higher-mass systems.  A new formulation based on
  pressure-regulated molecular cloud formation and on observations of
  low-pressure systems produces lower star formation rates and thus
  lower metallicities \citep{br06}.}  The zero-point offsets in the
left panels of Figure \ref{simulation} are probably a result of these
two effects.  We interpret these results as an encouraging sign that
the star formation and feedback models included in recent galaxy
formation simulations require only modest adjustments in order to
produce realistic stellar populations and chemical evolution.

In order to test whether the feedback in the simulation is what causes
the fundamental line, we also ran a set of simple dissipational
collapse simulations of disk galaxy formation.  These simulations are
comprised of 40,000 gas particles distributed in a fixed NFW dark
matter halo with a virial mass of $M_{\mathrm{vir}}=2.16 \times
10^{11} h^{-1}$~M$_{\sun}$ (the same as the galaxy in the cosmological
simulation from \citeauthor{robertson04} [2004]).  and a concentration
of 15.24 (appropriate for the mass; \citealt{bullock01}).  The initial
galaxy has a gas fraction $f_{\mathrm{gas}}=0.1$ and is given a spin
of $\lambda = 0.1$.  The simulations are then evolved for 2~Gyr.  In
each case, the gas cools and quickly settles into a disk (within
$\sim100$~Myr) and begins to form stars (see
\citealt{sh03,robertson04}).  We model the ISM in these simulations
using the \citet{sh03} multiphase model, modified to include an
effective equation of state parameter ($q_{\mathrm{EOS}}$) that
linearly interpolates between an isothermal gas ($q_{\mathrm{EOS}}=0$)
and an ISM strongly pressurized by supernova feedback
($q_{\mathrm{EOS}}=1$) as in the \citet{robertson04} simulation (see
\citealt{sh03}, \citealt{robertson04}, and \citealt{springel05} for
further details).  We simulate the dissipational collapse with various
values of $q_{\mathrm{EOS}}$ between 0.01 and 1.0 to study the impact
of the effects of supernova feedback on the fundamental line.

\begin{figure*}[th!]
\epsscale{1.20}
\plotone{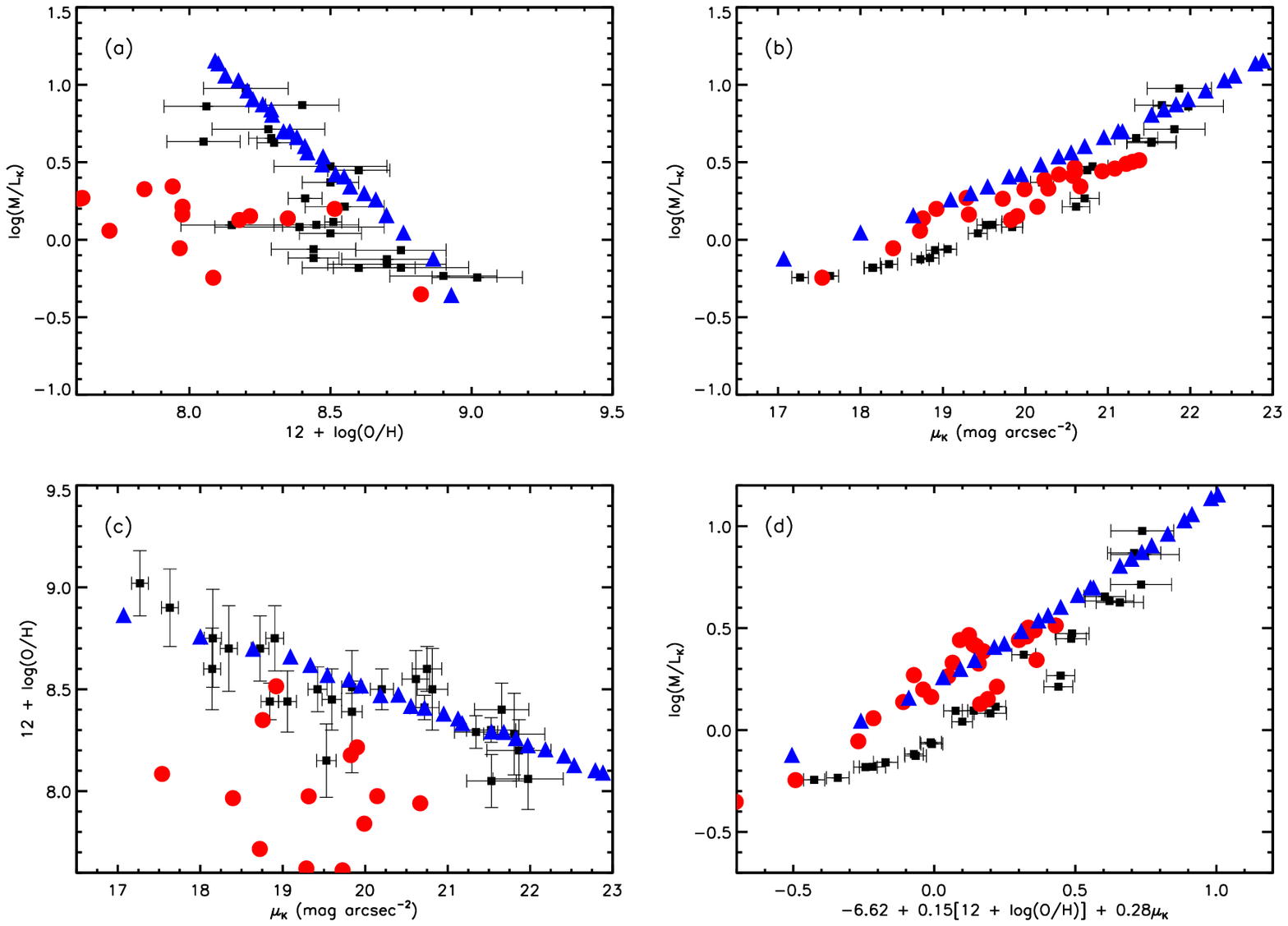}
\caption{Comparison of the scaling relationships in our dissipational
  collapse toy model simulations with the observed relationships
  in M33.  \emph{(a)} The M/L-metallicity relationship.  \emph{(b)}
  The M/L-surface brightness relationship.  \emph{(c)} The surface
  brightness-metallicity relationship.  \emph{(d)} The fundamental
  line.  The black squares are the M33 data points, and the red
  circles and blue triangles are the measurements from the simulations
  in the cases of weak ($q_{\mathrm{EOS}} = 0.01$) and strong
  ($q_{\mathrm{EOS}} = 1$) feedback, respectively.  The stronger
  feedback model clearly tightens all of the correlations and brings
  some of them closer to the observed relationships.
\label{feedback}}
\end{figure*}

We find that when $q_{\mathrm{EOS}}$ is close to 0 (i.e., feedback is
weak), all of the scaling relations have significant scatter (see
Figure \ref{feedback}).  In particular, for the M/L-metallicity and
surface brightness-metallicity relationships, the scatter is nearly
large enough to wipe out the correlations entirely.  There are also
noticeable zero-point offsets in each of the relationships, but since
these simulations are intended to be simple models rather than
realistic galaxies, we do not expect them to reproduce the observed
scaling relations with high fidelity.  Despite the scatter and the
failure to produce accurate metallicities, a fundamental line is still
apparent in these simulations.  As $q_{\mathrm{EOS}}$ is increased
towards 1, the relationships all tighten considerably, and the
simulated points converge on the real data.  This experiment suggests
that feedback is responsible for the tightness of the fundamental
line, but does not set the slope and zero point.  It is worth noting
that if we consider the M/L-metallicity, surface
brightness-metallicity, and fundamental line relationships in terms of
the mass-weighted gas+stellar metallicity instead of simply the ISM
metallicity, good correlations (with scatter) are obtained even in the
simulations with very weak feedback.  In fact, with this metallicity
prescription, the weak feedback correlations are generally closer to
the observed correlations than the strong feedback results are
(although again, they contain more scatter).  Since the gas and
stellar metallicities do not agree very well in these toy model
simulations, it is not clear which metallicity definition offers the
best comparison to our observations.

Although the results of these simulations appear promising in terms of
matching the fundamental line, the substantial offsets seen in some of
the component relationships suggest that we should be cautious in our
interpretation.  In a related example, while it is straightforward to
generate numerical galaxy models that match the slope of the
Tully-Fisher relation, simulations have not yet been able to
simultaneously reproduce the slope, scatter, and zero point of the
relation \citep[e.g.,][]{ns00,abadi03,dutton06}.  By analogy, it is possible
that the agreement we have found between the data and the simulations
is fortuitous.

\section{CONCLUSIONS}
\label{conclusions}

We have shown that there are strong correlations between dynamical
mass-to-light ratio and metallicity, dynamical mass-to-light ratio and
surface brightness, and metallicity and surface brightness that appear
to hold \emph{locally}, at every point within M33.  The trends of
these relationships are that low metallicity, low surface brightness
regions have high M/L, and high metallicity, high surface brightness
regions have low M/L.  By analogy with the fundamental line of dwarf
galaxies identified by PB02, we combine the three correlations into
the fundamental line of disk galaxies (Figure \ref{flplot}), using
local properties instead of global ones.  The M33 data follow the same
trends as the Local Group dwarf galaxies, but with zero-point offsets,
such that at a given metallicity, the mass-to-light ratio in M33 is
approximately a factor of $5-10$ higher than that of the dwarfs, and
at a given surface brightness M33 has a mass-to-light ratio that is 3
times higher than the dwarf galaxies.  These conclusions are largely
independent of the particular mass model chosen for M33.

We demonstrated using analytical calculations that the
metallicity-surface brightness relationship in galaxy disks is a
generic prediction of ordinary chemical evolution models.  Similarly,
the M/L-surface brightness correlation can be derived
straightforwardly under the assumptions of an exponential stellar disk
and a dark matter halo that follows any of the proposed structures in
the literature.  It is not yet clear whether the M/L metallicity
relationship shares a similar explanation or is merely a residual of
the other two correlations.  These results indicate that at least two
of the three relationships that make up the local fundamental line
originate from the basic properties of galaxy disks, and should
therefore be universally applicable to other galaxies.  Feedback,
which was proposed by PB02, \citet{dw03}, and \citet{tassis03} as the
physical mechanism responsible for the fundamental line of dwarf
galaxies, does not seem to play such an important role in the
fundamental line of disk galaxies.

Finally, we compared the fundamental line relationships in M33 with
their equivalents in a simulated disk galaxy from \citet{robertson04}.
We found good agreement in the slopes of each relationship, although
the metallicity of the simulated galaxy is too high.  Additional toy
model simulations showed that feedback controls the tightness of the
fundamental line, supporting the conclusion of our analytical models
that feedback is not necessary to create the local fundamental line.
Forcing simulated disks to match the fundamental line and its scatter
will improve current numerical recipes for star formation and
feedback.

\acknowledgements{We thank the anonymous referee for helpful comments.
  JDS gratefully acknowledges the support of a Millikan Fellowship
  provided by the California Institute of Technology.  FP is supported
  by the Spanish MEC under grant No. AYA2005-07789.  This research was
  also partially supported by NSF grant AST-0228963.  The simulations
  were performed at the Center for Parallel Astrophysical Computing at
  the Harvard-Smithsonian Center for Astrophysics.  We would like to
  thank Tom Jarrett for making the 2MASS data available to us prior to
  publication, and we acknowledge helpful discussions with Andrey
  Kravtsov.  This publication makes use of data products from the Two
  Micron All Sky Survey, which is a joint project of the University of
  Massachusetts and the Infrared Processing and Analysis
  Center/California Institute of Technology, funded by the National
  Aeronautics and Space Administration and the National Science
  Foundation.  This research has also made use of the NASA/IPAC
  Extragalactic Database (NED) which is operated by the Jet Propulsion
  Laboratory, California Institute of Technology, under contract with
  the National Aeronautics and Space Administration, and NASA's
  Astrophysics Data System Bibliographic Services.}

\end{document}